# Swimming statistics of cargo-loaded single bacteria

P. Prakash[1,a)], A. Z. Abdulla[2,b)], V. Singh[3], M. Varma[1,4,c]

[1]Centre for Nanoscience and Engineering, Indian Institute of Science, Bangalore, 560012, India

[2]Department of Physics, Indian Institute of Science, Bangalore, 560012, India

[3]Molecular Reproduction, Development and Genetics, Indian Institute of Science, Bangalore, 560012, India

[4]Robert Bosch Centre for Cyber Physical Systems, Indian Institute of Science, Bangalore, 560012, India

**Abstract**

Burgeoning interest in the area of bacteria-powered micro robotic systems prompted us to study the dynamics of cargo transport by single bacteria. In this paper, we have studied the swimming behaviour of oil-droplets attached as a cargo to the cell bodies of single bacteria. The oil-droplet loaded bacteria exhibit super-diffusive motion which is characterized by high degree of directional persistence. Interestingly, bacteria could navigate even when loaded with oil-droplets as large as 8 μm with an effective increase in rotational drag by more than 2 orders when compared to free bacteria. Further, the directional persistence of oil-droplet loaded bacteria was independent of the cargo size.

## 1. Introduction

The development of micro and nano-scale robots is emerging as an intensely active area of research motivated by their applications ranging from precise drug delivery to micro-scale manipulation.[1–3] It is envisioned that bacteria-powered bio-hybrid microbots will lead to self-powered, autonomous microrobotic systems for applications in targeted delivery.[4,5] Often these microbots derive their propulsive power and navigational control externally, for instance, from an applied magnetic field.[6] The internal generation of propulsive power at microscale has been a significant challenge prompting several groups to harness the propulsive power of bacteria in developing bio-hybrid robots at microscale.[7,8] In such bacteria powered bio-hybrid microbots, a cargo, such as a polystyrene bead is transported by collective swimming of several bacteria attached to the bead via chemical linkers.[9]

---

a) Present address: Warwick Integrative Synthetic Biology Centre, University of Warwick, Coventry, CV4 7AL, United Kingdom.
b) Present address: Laboratoire de Biologie et Modelisation de la Cellule, ENS de Lyon, 69364 Lyon Cedex 07, France.
c) **Author to whom correspondence should be addressed. Electronic mail: mvarma@iisc.ac.in**

Previous reports of cargo attachment on bacteria have relied on attaching cell body to very small particles (0.5 μm − 1 μm diameter) so that statistically only one cell binds due to steric restriction.[5,10] Similar chemical attachment process for larger cargo size ∼10 μm yields multiple bacteria 4 − 10 attached to a single cargo.[11] Such an approach cannot be utilised for studying the dynamics of a single cell carrying larger loads (∼10 μm diameter). Cargo carrying single bacteria is a preferred system to undertake studies as they are devoid of complex interactions present when multiple bacteria pushes a single cargo. In this work, we use a sonication-based method which allows us to attach cargo to a single bacteria in the form of an oil-droplet as large as 12 μm.[12] Using oil-droplet loaded on singular bacteria we were able to study the swimming statistics of cargo carrying bacterial trajectories.

## 2. Experimental

### 2.1. Attachment of oil-droplet cargos to single bacteria

We have used GFP (green fluorescent protein) labelled bacterial strain *Pseudomonas aeruginosa* (PA14) having two flagella at one end of the cell-body with average body length and diameter of $1.5 \pm 0.6$ μm and $0.6 \pm 0.2$ μm respectively in our study (Fig. 1(a)). A mixture of Silicone oil (Viscosity = 100 cst) and the buffer solution (0.5 mM PBS + 50 μM $MgCl_2.6H_2O$ + 10 μM EDTA) containing bacteria (∼$10^6$ cells/mL) in 1.5 mL eppendorf tube is sonicated for 30 – 45 seconds (Fig. 1(b)). Following sonication, the mixture is centrifuged to separate the aqueous phase containing bacteria from oil phase. This aqueous phase is then left idle for 8 − 9 hrs after which a fraction of bacteria in the solution is loaded with an oil-droplet, see Supplemental Material (SM) Section 1 for a detailed infographic. Due to the stochastic nature of the loading process, we observed three distinct groups of bacterial populations. There were free bacteria which were not loaded with oil-droplets, bacteria loaded with oil-droplet cargos of varying cargo size and bacteria which were rendered immobile due to large cargo size as shown in SM Video 1 and corresponding trajectories in SM Video 2 (see SM for description).

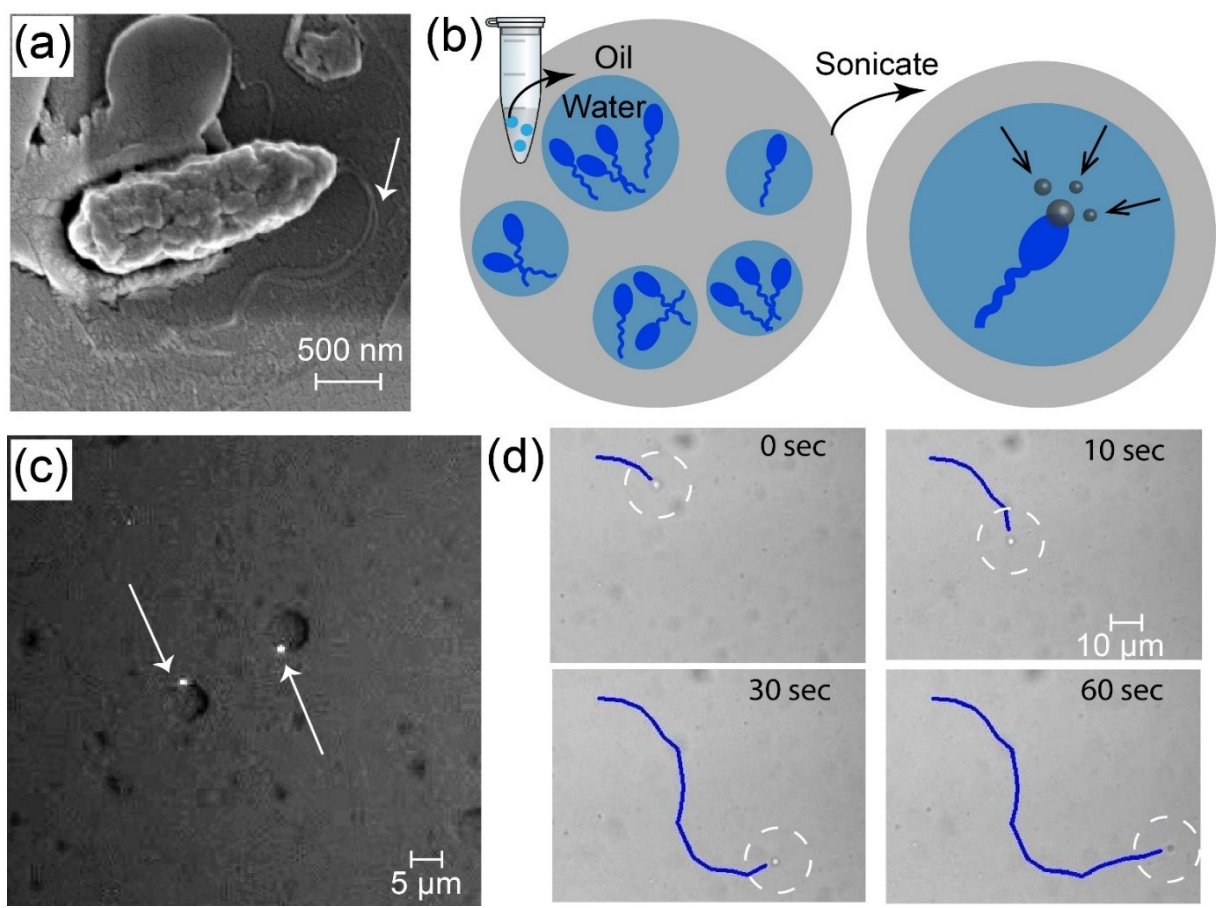

**Fig. 1** (a) Scanning electron microscopy of *Pseudomonas aeruginosa* (PA14) reveals an average length of 1.5 μm and width 0.6 μm with two flagella protruding from one end. (b) Schematic representation of loading oil-droplet on bacteria. 10 μl of bacterial solution and 200 μl of Silicone oil is mixed and sonicated to load oil-droplet on bacteria. (c) Fluorescence microscopy showing bacteria embedded onto oil-droplet. (d) Track of oil-droplet loaded on bacteria.

## 2.2 Microscopic imaging of cargo-loaded bacteria

The aqueous phase after maturation period of 8-9 hours is sandwiched between glass slide and cover slip separated by sticky double tape which provides a fluidic gap of ∼200 μm. Concurrent imaging is performed using Differential Interference Contrast (DIC) and Fluorescence (FL) filters by manually switching between the two modes. This allowed us to easily distinguish the GFP labelled bacterial body from the oil-droplet under the microscope as shown in Fig. 1(c). We recorded several oil-droplet loaded bacteria trajectories for further analysis (see SM Video 3). The oil-droplet loaded bacteria can comfortably swim laterally as well as along the thickness of the flow cell as evident from the variation in the focus of the oil-droplet as shown in Fig. 1(d). The movies were captured at a magnification of 20X in the DIC mode after confirmation of bacterial attachment in the Fluorescent mode.[13]

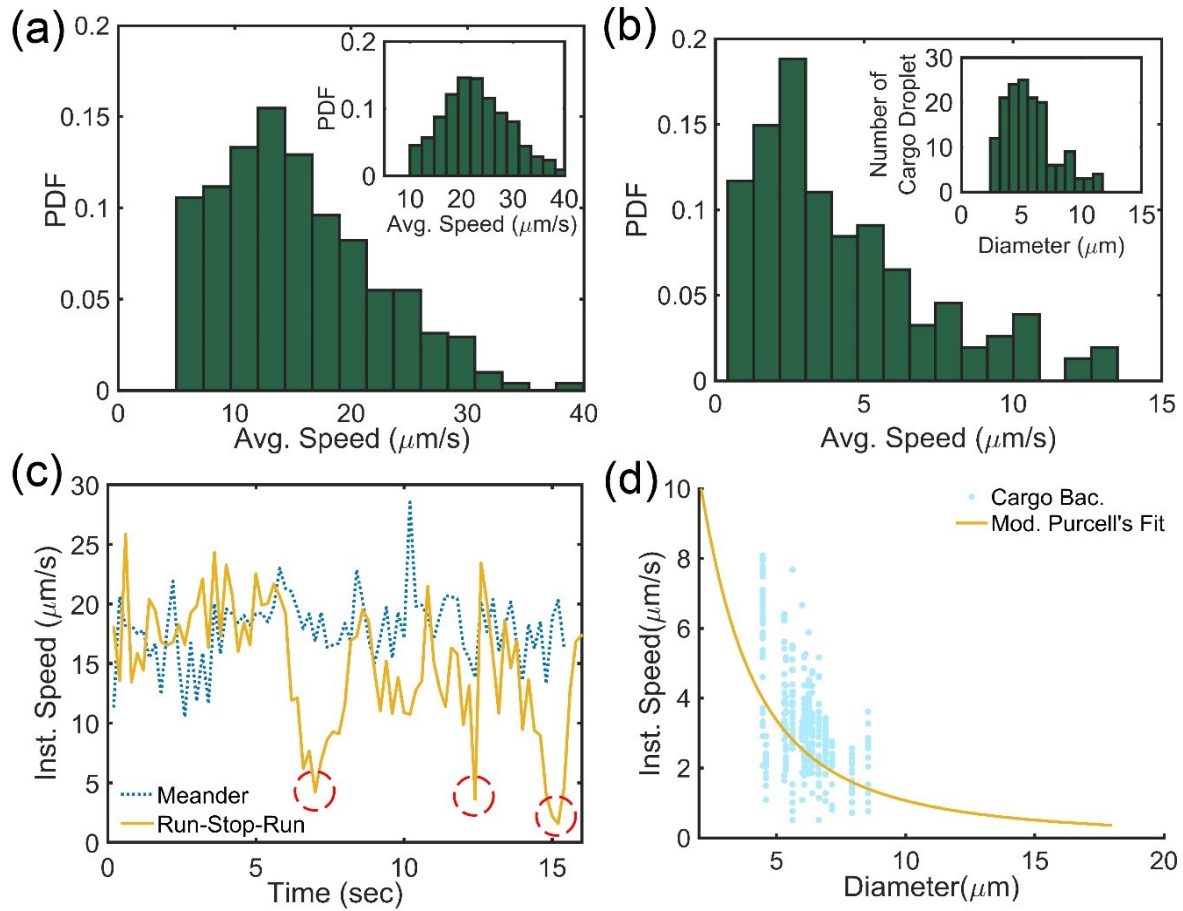

**Fig. 2** (a) The average speed of starved bacteria *Pseudomonas aeruginosa* over 511 tracks is 15.3 μm/sec. The inset shows average speed histogram of fresh bacteria over 1814 tracks. (b) The average speed of oil-droplet loaded bacteria over 154 tracks drops to 4.3 μm/s. The inset image shows histogram of the size of oil-droplets, where average size is 5.75 μm. (c) Bacteria exhibits meander and run-stop-run events, where, the stop events are marked when instantaneous speed drops below 5 μm/s. (d) The speed of cargo-loaded bacteria vs. size can be fit using a modified form of Purcell's model.

## 3. Results

### 3.1 Swimming speed statistics

The average swim speed of the native free bacteria taken directly from the growth medium was found to be around ∼23 μm/s (inset Fig. 2(a)). This is similar to the reported average swim-speed of wild-type strain of, *E.coli* bacteria, i.e. around ∼20 μm/s.[14,15] Further, as described in Section 2, after the cargo attachment procedure, there is still a fraction of bacterial population which are not loaded with oil-droplet cargo. We observed that the average swim-speed of these free bacteria which do not have oil-droplet cargo attached to them drops to ∼15 μm/s (Fig. 2(a)). The reason for this drop in the average swim-speed is probably due to nutrient deprivation during approximately $8-9$ hour long maturation process required for the growth of oil-droplets. The bacteria loaded with oil-droplet cargos swim even slower. The swim-speed of cargo-loaded bacteria averaged over 154 tracks covering cargo diameters ranging from 2.5 µm − 12 µm was 4.3 µm/s (Fig. 2(b)). The cut-off at lower average speed of motility data is due to the software limitations in correctly tracking bacteria at lower speed. Similarly, the identification of smaller sized oil-droplet is limited by the magnification of microscope objective which is 20X in this case. Interestingly, despite the large relative size of the oil-droplet cargo, we found this system to be quite stable and did not observed any instance of cargo detachment in our experiments. The hydrophobic nature of the oil-droplet cargo and bacteria cell-membrane may account for the observed stability during the experiments.[16,17]

Though, *P. aeruginosa* is widely used as a model system of polar mono-flagellated bacterial species; ours is a bi-flagellated species as revealed from the Scanning Electron Microscopy (SEM) image of our strain (Fig. 2(a), see SM Section 2 for raw images). Various strains of *P. aeruginosa* are indeed reported to have more than one flagella.[18,19] To further explore the motility behaviour, we analysed instantaneous speed of our strain as shown in Fig. 2(c). Interestingly, the tracks exhibit meander as well as run-stop-run kind of motion, where, speed drops drastically during the 'stop' events.[20] The sudden drop in speed of bacterial species having more than one flagella such as *E. coli* are known to be signature of tumbling events.[21] As oil-droplet loading process involves of unconventional steps such as sonication and nutrient deprivation, we verified, if the loaded bacteria indeed display wild-type motility behaviour. For analysing their motility, we used a modified form of Purcell's model to include extra drag because of oil-droplet cargo as described in our recent report – PRE, **100**, 2019, 062609. The speed of oil-droplet loaded bacteria is inversely proportional to the size of cargo and fits well to the modified Purcell's model as shown in Fig. 2(d).

### 3.2. Directional persistence

The swimming bacteria performs ballistic motion for short time-scales which eventually transitions to diffusive motion for longer times.[22] Consequently, the mean-squared displacement $\left(\text{MSD} =< \left(r(t+dt) - r(t)\right)^2 >\right)$ of bacteria scales as $(dt)^2$ for short time-scales and as $(dt)^1$ for longer time-scales characterizing ballistic and diffusive behaviours respectively. The MSD of free-swimming PA14 (averaged over 41 trajectories; tracks shown Section 3 of the SM) shown in Fig. 3(a) exhibits a transition from ballistic (line with slope = 2) to diffusive behaviour (line with slope = 1) in 3.6 sec. The mean squared displacement

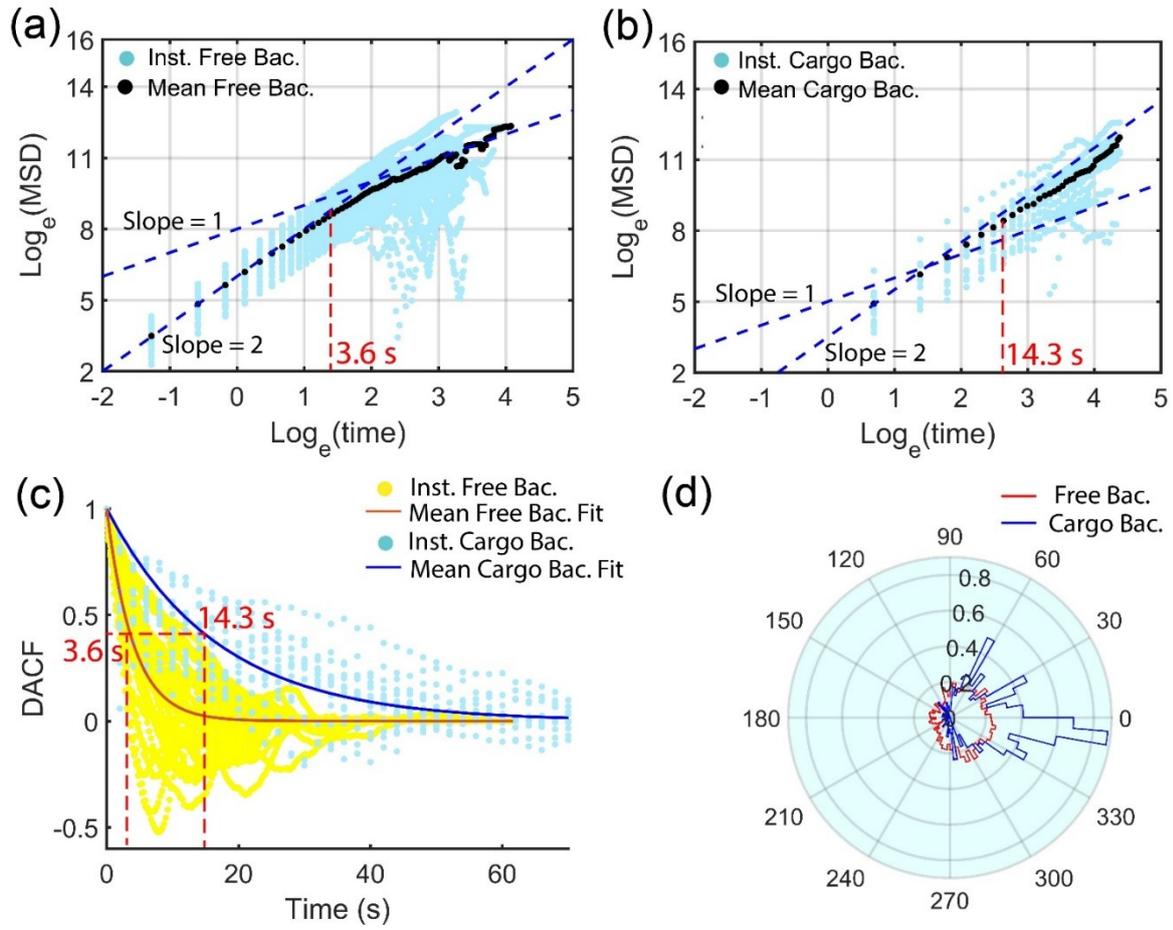

**Fig. 3** Motility analysis of free and cargo-loaded bacteria. (a) The MSD plot of bacteria confirms an initial ballistic motion (slope = 2 on log-scale plot) which eventually becomes diffusive (slope = 1) due to tumbling events. (b) The cargo-loaded bacteria exhibit super diffusive behavior (slope ~ 1.8). (c) Directional autocorrelation function shows an increased persistence in the motion of cargo loaded bacteria (cyan) as compared to free bacteria (yellow). (d) Polar plot of the distribution function of bacterial orientation relative to initial direction. Cargo-loaded bacteria proceed close to their initial direction indicating high persistence in swim direction.

(MSD) for diffusive motion scales with time as $Log(MSD(t)) = Log(6D) + Log(t)$. Accordingly, the translational diffusion coefficient of bacteria is estimated as 497 μm²/s from the intercept of Log (MSD) vs Log (t) curve using the diffusive regime (line with slope = 1) (Fig. 3(a)). The transition of bacteria from ballistic to diffusive motion can also be estimated independently from the direction auto-correlation function DACF, defined as $DACF(\tau) = < \hat{n}(t) \cdot \hat{n}(t+dt) >$, where $\hat{n}(t)$ is the unit vector in the swimming direction at time t. To extract the re-orientation time or run time ($\tau_R$) of swimming bacteria, we calculated the direction auto-correlation function and fit it to the exponential function $e^{-t/\tau_R}$ to obtain $\tau_R = 3.6$ sec as shown in Fig. 3(c).[23,24] One can use the reorientation time-scale to estimate an effective diffusion coefficient $D_{eff}$ of bacteria swimming with an average speed of $V_0$, as $D_{eff} =$

$V_0^2 \tau_R / 3(1-\alpha)$.[23] Here, α is the mean of cosine of the re-orientation angle. The re-orientation angles of free bacteria were extracted yielding the value of α as 0.35 (see Sec. 2 of the SM). Using the measured average speed of free bacteria ~15 μm/s, we obtain a diffusion constant of 415 μm²/s which is reasonably close to 497 μm²/s extracted from the MSD plot. In contrast to the free bacteria, the MSD plot of cargo-loaded bacteria do not transition to diffusive behaviour within the maximum observation window of around 90 seconds. Instead, the MSD of cargo-loaded bacteria (averaged over 16 trajectories, tracks shown in SM Section 3) exhibits super diffusive behaviour with a slope of 1.8 (Fig. 3(b)). Though, we cannot obtain a diffusion coefficient in the same manner as we obtained for the case of free bacteria. We can still obtain a reorientation time scale for cargo-loaded bacteria based on the DACF, which turns out to be 14.3 seconds (Fig. 3(c)). It can be noticed in the MSD plot, that this corresponds to the timescale at which the experimental data starts deviating from the ballistic regime (line with slope = 2). The longer reorientation time of cargo-loaded bacteria translates to increased directional persistence resulting in super diffusive dynamics. This is more clearly seen when we plot the polar probability distribution function of deviation $(\theta(\tau_i) - \theta(0))$, from the initial direction $\theta(0)$ where, $\tau_i$ is swept over entire track duration (Fig. 3(d)). For the oil-droplet loaded bacteria we observe a prominent peak around initial direction as compared to more uniformly distributed deviation in the case of free bacteria. The increased directional persistence can affect the optimal strategy for navigation in different geometry.[25–27] For instance, free bacteria exhibit diffusive motion whereas the same bacteria in swarming state, translocate in a super-diffusive manner similar to the cargo-loaded bacteria described here.[28]

### 3.3 Angular deviation as a function of cargo size.

We analysed the angular deviation of bacteria for τ = 0 − 10 sec and cargo loaded bacteria in τ = 0 − 50 sec averaged over all tracks $\theta_d(\tau) = <\theta(t + \tau) - \theta(t)>$ as shown in Fig. 4(a). The time axis is normalized by the reorientation time for the angular deviation plot of bacteria $\tau_R = 3.6$ sec and cargo loaded bacteria $\tau_R = 14.3$ sec respectively. The normalization of time allows us to plot the angular deviation of entire 10 sec and 50 sec bacteria and cargo loaded bacteria trajectories in one plot. The average angular deviation of bacterial tracks is 60° in a time interval of 12.5 sec (roughly 3.5$\tau_R$, $\tau_R = 14.3$ sec ) whereas it is only around 30° in 50 sec (roughly 3.5, $\tau_R = 14.3$ sec) for oil-droplet loaded tracks (Fig. 4(a)). Clearly, the large angular deviation in bacteria leads to their transit from ballistic to diffusive behaviour which is not observed in oil-droplet loaded bacteria tracks. We then analysed the dependence of

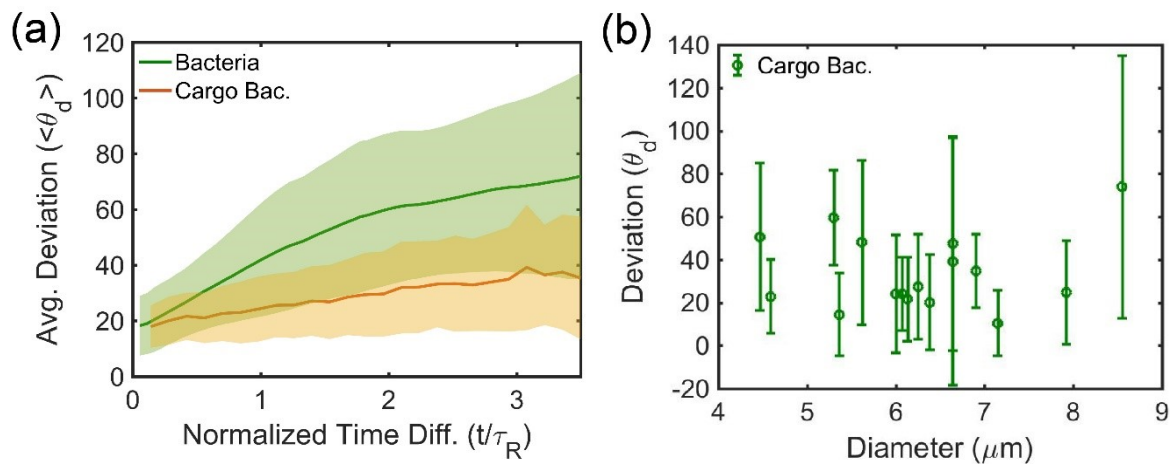

**Fig. 4** (a) Angular deviation of bacteria and oil-droplet loaded tracks averaged over all tracks. The shaded area is the standard error of angular deviation. (b) Angular deviation of oil-droplet loaded tracks after 40 sec.

directional persistence on the size of the cargo. Due to the stochastic nature of oil-droplet size distribution, it is not possible to generate large number of trajectories for a given cargo size. However, as the tracks are over 50 sec long, we get averaged data for at least 10 values per track. Since bacteria performs random walk, an average deviation over several equidistant segments for one track is statistically as significant as average over several trajectories. The angular deviation ($\theta_d$) of individual tracks after 40 sec for oil-droplet loaded tracks as a function of their size is shown in Fig. 4(b). Surprisingly, angular deviation shows no significant dependence on the size of cargo.

### 3.4 Can we explain angular deviation by thermal or tumbling reorientations?

A change in direction of a multi-flagellated bacteria can occur either due to tumbling events or thermal diffusion. A tumbling event is distinctly identified by a sudden drop in speed whereas, the reorientation due to thermal diffusion can be estimated analytically.[29] From the direct measurements of speed, we see that our strain exhibits meander as well as run-stop-run motion as shown in Fig. 2(c). The stop events in the case of free PA14 is adjudged whenever there is a dip of more than 3 times in the average speed.[20] Video evidence of tumbling event can be clearly seen in SM Video 1 where, several bacterial trajectories slow down to change their direction. In particular, sharp directional change (hallmark of tumbling events) can be noticed in Trajectory 13 (time stamp 40 sec onwards) of the corresponding tracked movie SM Video 2. Out of 41 free PA14 trajectories, 13 tracks underwent 2-5 stop events whereas 28 trajectories underwent either 0 or 1 stop events meaning nearly ~68% of the total tracks are meander trajectories. The time difference between consecutive stops averaged over 13 trajectories is

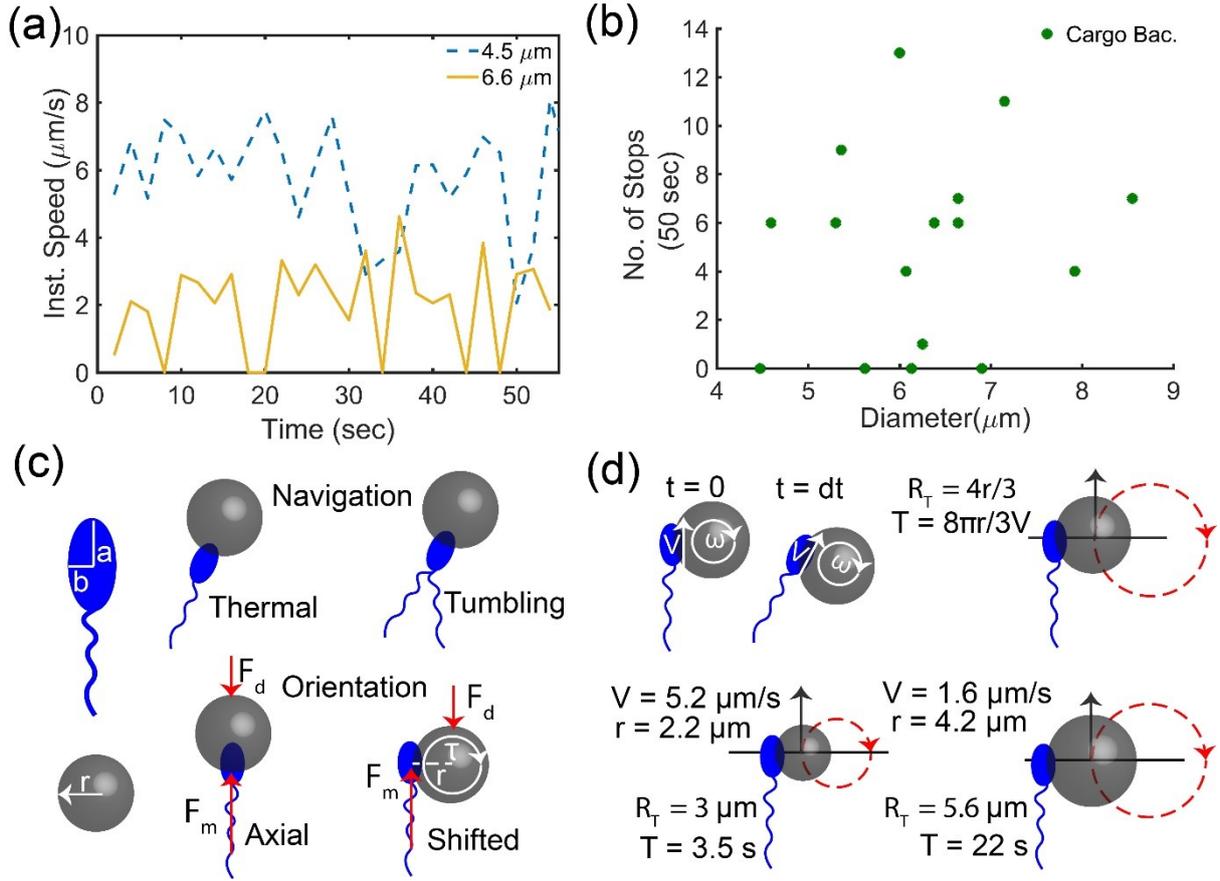

**Fig. 5** (a) Cargo loaded bacteria exhibiting meander and run-stop-run events. (b) Number of stops vs Cargo size in 50 sec. (c) A schematic showing thermal or tumbling mode of directional change. The oil-droplet attachment process is stochastic so it can attach to any part of bacterial body. (d) If the oil-droplet is attached on the bacterial body from sideways, then it is constrained to move in circular trajectory.

~ 5 sec which is close to the reorientation timescale of 3.6 seconds obtained from experiments Fig. 3(a). Similarly, the orientation of is also affected by rotational diffusion due to thermal motion. The rotational diffusion constant for free bacteria in thermal equilibrium is $D_{rot} = \frac{kT}{D_o}$ where, T = 300 K, drag coefficient $D_o = \frac{8\pi\eta_w a^3}{3(\ln(2a/b) - 1/2)}$.[30] Considering the free bacteria as a prolate spheroid as shown in Fig. 6(c) with a = 0.75 μm and b = 0.3 μm in aqueous solution with viscosity $\eta_w$. We get $D_{rot} = 0.3$ rad$^2$/s from which the reorientation timescale can be estimated as $\tau_R = 1/2D_{rot} = 1.6$ sec which is of the similar order of experimentally observed value of 3.6 sec. The difference in reorientation timescale is within the experimental error, as the estimate is highly sensitive due to cubic dependence of rotational drag $D_o$ on the cell body length (2a = 1.5 ± 0.6 μm). So, if we change a from 0.74 μm to 0.94 μm then the reorientation time will change by a factor of 2.

Similarly, the oil-droplet loaded trajectories also display meander and run-stop-run trajectories as shown in Fig. 5(a). Out of 16 trajectories analysed, 4 trajectories performed run-stop-run motion where, the average number of stops is ~7 in 50 sec. The average time consumed per stop in the case of cargo loaded bacteria has nearly doubled and is reminiscent of decrease in tumbling frequency of bacteria observed in polymeric medium.[20] The experimentally obtained reorientation time for the cargo-loaded bacteria is ~14.3 sec which is well within the frequency of stops ~ 7 sec in the oil-droplet loaded bacteria. However, tumbling alone can't describe the directional change as nearly 75% of the tracks do not exhibit stop events at all. Moreover, we see no discernible pattern in the number of stops vs size of cargo as shown in Fig. 5(b). As before, the reorientation time due to thermal diffusion of oil-droplet loaded bacteria is $1/2D_{rot}$, where $D_{rot} = \frac{kT}{P} = 9.2 \times 10^{-3}$ rad$^2$/s, T = 300 K, drag coefficient $P = 8\pi\eta_w r^3$, average radius of oil-droplet r = 6.25 μm. As the size of cargo is much bigger than the size of bacteria, the torque required for rotation is predominantly determined by the rotational drag on oil-droplet. The reorientation time scale from thermal diffusion comes out to be 55 sec which is much larger than the experimentally observed value of 14.3 sec. Hence, we can't explain the observed independence in angular deviation as shown in Fig. 4(b) from directional fluctuations due to thermal diffusion.

## 3.5 Trajectory of oil-droplet loaded bacteria

A bacteria is propelled by rotating helical flagella which couples the rotational and translational degree of freedom pushing the body forward.[31] In contrast, bacterial body and oil-droplet are symmetric so the transverse rotational motion of bacterial body doesn't affect its translational motion. If the oil-droplet is attached to the head of bacteria, then it merely increases the rotational drag without having any effect on trajectory. However, if the droplet is attached on the bacterial body from sideways as shown in Fig. 5(c), then it can alter the course of trajectory. In such scenario, as the bacteria moves forward the oil-droplet rotates in forward direction as shown in schematic Fig. 5(d). As the bacteria moves forward with velocity 'V', the angular velocity ω of the oil-droplet in forward direction can be calculated from:

$$\tau = 8\pi\eta_w r^3 \omega = r \times F_m = r \times 6\pi\eta_w rV$$

where, r is the radius of oil-droplet, $F_m = 6\pi\eta_w rV$ is the translational force generated by bacteria and $\eta_w$ is the viscosity of the aqueous medium. Since the bacteria is always constrained to rotate on oil-droplet, the resulting trajectory will be circular. Assuming, the radius of trajectory of the centre of the oil-droplet to be $R_T$, for time 'dt' we can write:

$$dt = R_T d\theta/V$$

By replacing $d\theta/dt = \omega = 3V/4r$, we get the radius of trajectory as $4r/3$ and the corresponding time period $T = 8\pi r/3V$ for completing one orbit. Therefore, if the bacterial body is attached sideways, then its motion will be constrained to move in a circular trajectory with radius 1.33 times the radius of cargo. This may seem surprising, however, because of the dynamical constraints ($\tau = r \times F_m$) an increase in velocity also results in corresponding increase angular velocity $\omega$ leading to faster rotation. Hence, an increase in velocity simply results in faster rotation decreasing the time period. In our case the smallest oil-droplet is of the size 4.4 μm and moves with a speed of 5.2 μm/s; assuming the bacteria to be attached sideways the resulting trajectory will be of radius 3 μm. Similarly, the largest oil-droplet is of size 8.4 μm and moves with a speed of 1.6 μm/s which if attached sideways, would result in a trajectory of radius 5.6 μm as shown in Fig. 5(d). We have seen evidences of such trajectories in our videos, see time stamp 5 − 30 sec in SM Video 1 and corresponding track in SM Video 2. Clearly a sideways attachment of oil-droplet on bacterial body won't lead to long trajectories as observed (see, SM Section 3). Nevertheless, the time period calculation of such trajectories reveals that it would barely take 0.3 sec for the smallest and 1.8 sec for the largest sized bacteria to turn by an angle of 30°. This means even a slight obstruction can result in toque generation which can quickly rotate the oil-droplet loaded bacteria. We indeed see oil-droplet attached to bacteria quickly change their direction in the presence of stationary debris in the form of oil-droplets (see, several trajectories in SM Video 3). This may lead to sustained angular deviation (Fig. 4(b)) even for larger sized oil-droplets as they have higher probability of encountering debris in a uniformly littered environment.

## 4. Conclusion

To summarize, we show that the oil-droplet loaded bacteria exhibits super diffusive motion with high directional persistence compared to the free bacteria. Interestingly, we did not observe any dependence of the cargo-size on the directional persistence. This may be explained by the directional change which occurs as the moving oil-droplet loaded bacteria encounters other stationary oil-droplets present in the system. Further experimental and corresponding theoretical analysis is warranted to understand the role of debris in the navigational capability of cargo loaded bacteria systems.

# Supplemental Material

## Swimming statistics of cargo-loaded single bacteria

Praneet Prakash,[a,b] Amith Z. Abdulla,[c,d] Varsha Singh,[e] and Manoj Varma [a,f*]

[a]Centre for Nanoscience and Engineering, Indian Institute of Science, Bangalore, 560012, India
[b]Present address: Warwick Integrative Synthetic Biology Centre, University of Warwick, Coventry, CV4 7AL, UK.
[c]Department of Physics, Indian Institute of Science, Bangalore, 560012, India
[d]Present address: Laboratoire de Biologie et Modelisation de la Cellule, ENS de Lyon, 69364 Lyon Cedex 07, France.
[e]Molecular Reproduction, Development and Genetics, Indian Institute of Science, Bangalore, 560012, India
[f]Robert Bosch Centre for Cyber Physical Systems, Indian Institute of Science, Bangalore, 560012, India
[*]**Author to whom correspondence should be addressed. Electronic mail: mvarma@iisc.ac.in**

### 1. Infographic on oil-droplet attachment process

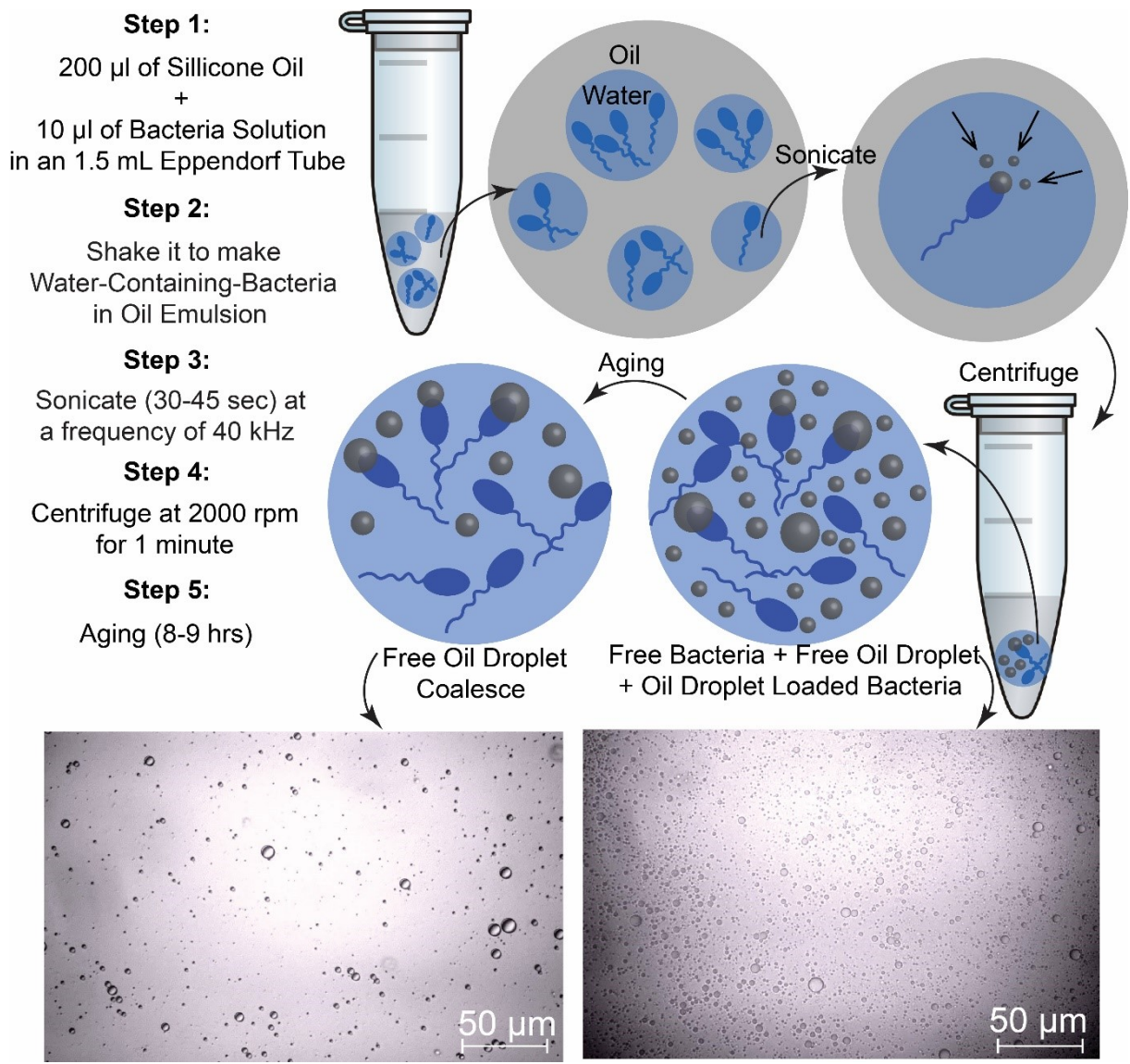

## 2. Raw Scanning Electron Microscopy image of bacteria

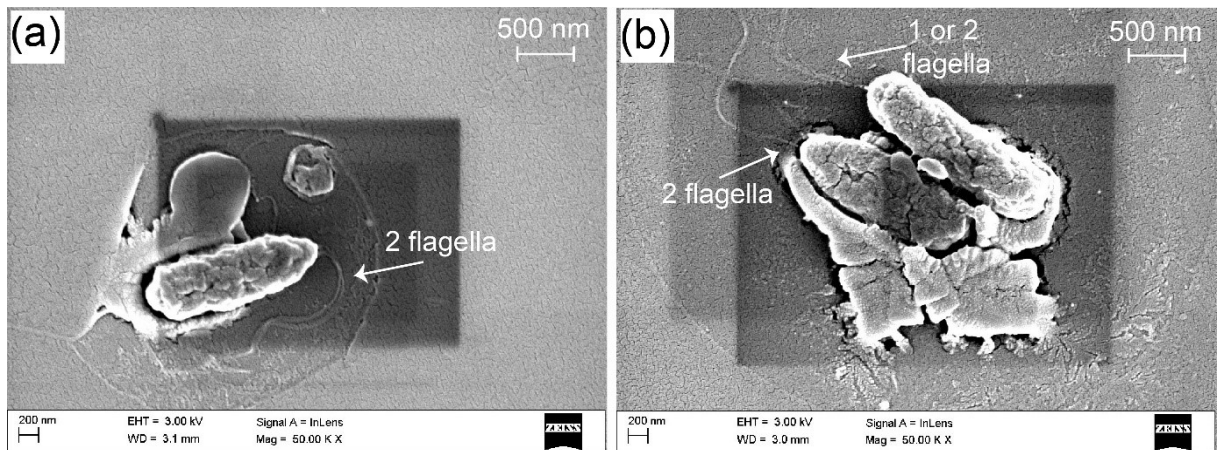

**Fig. S1** Raw SEM image showing 2 flagella protruding from two different bacteria. To our eye, two flagella seems to be coming of even from the third bacteria but is ambiguous. As we have observed several meandering tracks it may be that this strain is a mixture of bi-flagellated and mono-flagellated strain.

## 3. Free and oil-droplet loaded bacteria tracks

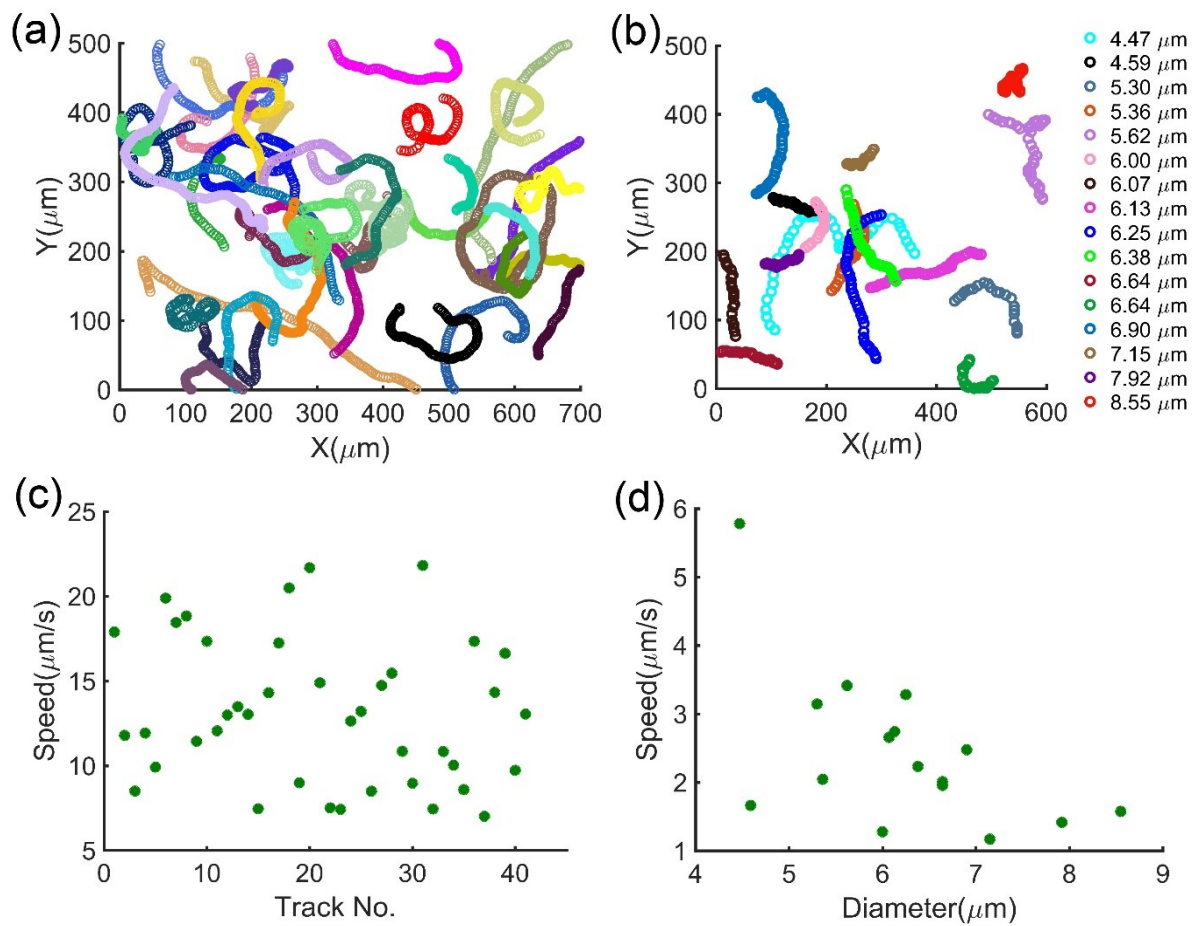

**Fig. S2** (a) Bacteria tracks: 35 tracks are up to 25 sec long and 5 tracks are up to 40 sec long (b) Oil-droplet loaded bacteria track: 8 tracks are up to 50 sec long and another 8 tracks are up to 80 sec. (c) Average speed of bacteria tracks. (d) Average speed of oil-droplet loaded bacteria tracks.

## 4. $\alpha = <\text{Cos}(\theta_d)>$ for bacteria and oil-droplet loaded bacteria tracks

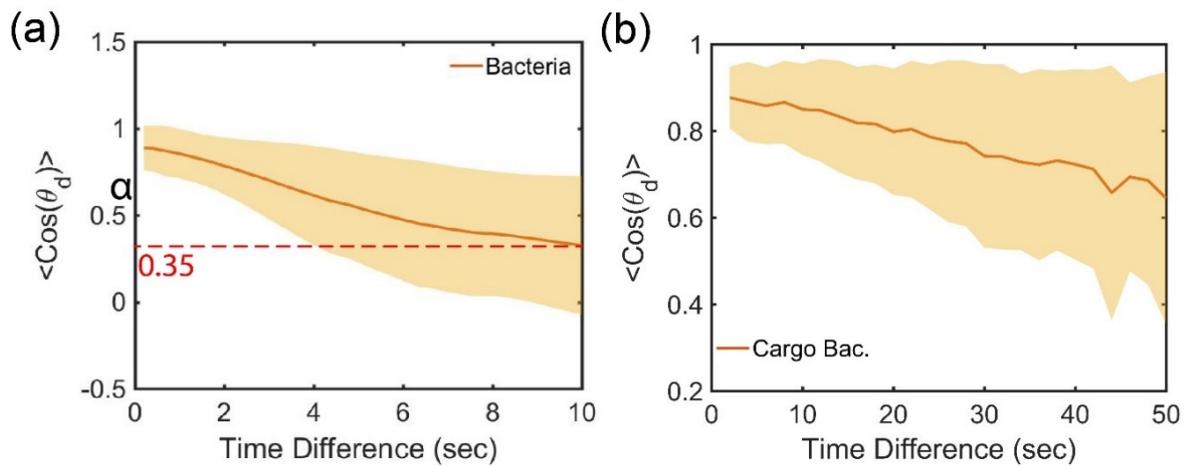

**Figure S3:** (a) $\alpha = 0.35$ for bacteria tracks after 10 sec. (b) $\alpha$ for oil-droplet loaded bacteria tracks.

## 5. Description of the Supplementary Videos

Video 1 shows following four types of motile species:

a) Several motile bacteria throughout the movie; corresponds to Track No. 4-17 in Video 2.

b) Oil-droplet loaded bacteria $(0 - 4\ \text{sec})$; corresponds to Track No. 1 in Video 2.

c) Oil-droplet attached sideways to bacteria $(5 - 30\ \text{sec})$; corresponds to Track No. 2 in Video 2.

d) Oil-droplet attached to a large oil-droplet which is unable to move (30 sec onwards); corresponds to Track No. 3 in Video 2.